\documentstyle[amssymb,aps,twocolumn,epsf]{revtex}

\begin{document}
\draft
\title{Resistivity and $1/f$ Noise in Non-Metallic Phase Separated
Manganites}
\author{A.~L.~Rakhmanov$^{a}$, K.~I.~Kugel$^{a}$, Ya.~M.~Blanter$^{b,c}$,
and
M.~Yu.~Kagan$^{d,e}$}
\address{$^a$ Institute for Theoretical and Applied Electrodynamics,
Russian Academy of Sciences, 13/19 Izhorskaya Str., 127412 Moscow,
Russia\\
$^b$ D\'epartement de Physique Th\'eorique, Universit\'e de
Gen\`eve, CH-1211 Gen\`eve 4, Switzerland\\
$^c$ Department of Applied Physics and DIMES, Delft University of
Technology, Lorentzweg 1, 2628 CJ Delft, The Netherlands\\
$^d$ Kapitza Institute for Physical Problems, Russian Academy of
Sciences, 2 Kosygina Str., 117334 Moscow, Russia\\
$^e$ Max-Planck-Institut f\"ur Physik Komplexer Systeme, N\"othnitzer
Strasse 38, D-01187 Dresden, Germany}
\date{\today}
\maketitle

\begin{abstract}
A simple model is proposed to calculate resistivity, magnetoresistance, and
noise spectrum in non-metallic phase-separated manganites containing small
metallic droplets (magnetic polarons). The system is taken to be far from
the percolation transition into a metallic state. It is assumed that the
charge transfer occurs due to electron tunneling from one droplet to another
through the insulating medium. As a result of this tunneling, the droplets
acquire or lose extra electrons forming metastable two-electron and empty
states. In the framework of this model, explicit expressions for dc
conductivity and noise power of the system are derived. It is shown that the
noise spectrum has $1/f$ form in the low-frequency range.
\end{abstract}


\bigskip \tighten

\section{Introduction}

Recent experimental \cite
{Allodi,Hennion,Morimoto,Babushkina,Voloshin,Podzorov} and theoretical \cite
{Nagaev,Gorkov,Arovas,Dagotto,KKM,KKK} papers provide a strong evidence for
the existence of the phase separated state in the perovskite manganites with
the colossal magnetoresistance (CMR). Both experiment and theory demonstrate
that under certain conditions the material in the phase separated state
consists of small ferromagnetic metallic droplets, or magnetic polarons,
embedded into the insulating antiferromagnetic (AFM) matrix. The droplet in
the ground state contains one charge carrier (electron or hole) confined in
a potential well of ferromagnetically ordered local spins. In Refs. %
\onlinecite{KKM,KKK} the model allowing the estimation of the droplet radius
and droplet concentration was proposed. According to this model, the radius
of the droplet $a$ is defined from the minimization of the energy: $E \sim
t( \pi d/a)^2+JS^2 (4 \pi /3) (a/d)^3$, where the first term is related to
the kinetic energy of electron delocalization in the bubble with the radius
$%
a$, and the second term corresponds to the loss in the energy of the
Heisenberg AFM exchange due to ferromagnetic ordering of local spins $S$
inside the bubble. The minimization of the energy with respect to droplet
radius yields the following estimate for $a$ in $3D$ case: $(a/d) \sim
(t/JS^2)^{1/5}$, where $J$ is the AFM Heisenberg exchange, $t$ is the
bandwidth, and $d$ is the intersite distance. The number of charge carriers
is proportional to the electron (hole) doping $x$. Volume concentration of
metallic droplets increases with $x$ and with the decrease of the
temperature since the droplet radius $a$ decreases with temperature growth.
As a result, at some critical concentration of holes $x_c \sim (d/a)^3$ the
droplets start to overlap and the percolation metal-insulator transition
occurs in the system \cite{Gorkov}. In Ref. \onlinecite{Podzorov} it was
observed that heating and cooling of the perovskite manganites in the phase
separated state is accompanied by strong hysteresis in the magnetization and
the resistance. In addition, the giant $1/f$ noise was measured in these
experiments. The noise power is very high even far from the percolation
threshold and drastically increases in its vicinity. The noise spectrum is
close to $1/f$ form in the $1$ -- $1000$ Hz frequency range.

In this paper, we calculate conductivity, magnetoresistance, and noise
spectral power of the system in the phase separated state. The calculations
are based on the results of Refs. \cite{KKM,KKK} and on a simple model for
the tunneling conductivity of the material accounting for electron jumps
from one polaron to another. The concentration range not too close to the
percolation transition is considered.

We use the terms {\em (magnetic) polarons} and {\em droplets}
interchangeably throughout the paper.

\section{Conductivity}

\label{condsect}

Let us consider an insulating antiferromagnetic sample of volume $V_s$ in
electric field $\bbox{E}$. The total number of magnetic polarons in the
volume is $N$, and thus their spatial density is $n = N/V_s$. As it was
mentioned before, the number of polarons is assumed to be equal to the
number of charge carriers introduced by doping. Neglecting the conductivity
of the insulating phase, we assume that charge carriers are only located
within the droplets. The charge transfer can thus occur either due to the
motion of the droplets or due to the electron tunneling. The former
mechanism is less effective: Indeed, the motion of a droplet is accompanied
by a considerable rearrangement of the local magnetic structure, which
results in the big effective mass of magnetic polarons. In addition, the
droplets are expected to be easily pinned by crystal lattice defects. Thus,
it is realistic to assume that the charge transport is essentially due to
electron transitions between the droplets.

A magnetic polaron in the ground state contains one electron. As a result of
a tunneling process, droplets with more than one electron are created, and
some droplets become empty (the lifetime of such excitations is discussed in
the end of this Section). If the energy of an empty droplet $E(0)$ is taken
to be zero, the energy of a droplet with one electron can be estimated as $%
E(1) \sim t(d/a)^2$. This is essentially the kinetic energy of an electron
localized in the sphere of the radius $a$. In the same way, the energy of
two-electron magnetic polaron $E(2) \sim 2E(1) + U$, with $U$ being the
interaction energy of the two electrons. In all these estimates, we have
disregarded the surface energy, which is expected to be small \cite{KKM}.
Thus, $E(2) + E(0) > 2E(1)$, and the creation of two-electron droplets is
associated with the energy barrier of the order of $A \equiv E(2) - 2E(1)
\sim
U $. It is clear that the interaction energy $U$ of two electrons in one
droplet is determined mainly by the Coulomb repulsion of these electrons,
hence $A \sim e^2/\epsilon a$, where $\epsilon$ is the static dielectric
constant, which in real manganites can be rather large ($\epsilon \sim 20$).
We assume below that the mean distance between the droplets is $n^{-1/3} \gg
a$ (the droplets do not overlap). Then, $A$ is larger than the average
Coulomb energy $e^{2}n^{1/3}/\epsilon$. Since the characteristic value of
the droplet radius $a$ is of the order of 10 \AA \cite{KKM,KKK}, we have $%
A/k_B \sim 1000 K$ and $A>k_{B}T$ in the case under study. In the following,
we assume that the temperature is low, $A \gg k_BT$, and we do not consider
a possibility of the formation of the droplets with three or more electrons.
Even in the case when these excitations are stable, it can be shown that far
from the percolation threshold the strong Coulomb interaction suppresses
their contribution to the conductivity (giving rise only to the next order
terms with respect to $\exp(-A/k_BT)$).

Let us denote the numbers of single-electron, two-electron, and empty
droplets as $N_1$, $N_2$, and $N_3$, respectively. According to our model,
$%
N_2 = N_3$, $N_1 + 2N_2 = N$, and $N$ is constant. Before turning to
conductivity, we evaluate the thermal averages of $N_1$ and $N_2$. To this
end, we note that the number $P_N^m$ of possible states with $m$
two-electron droplets and $m$ empty droplets equals $C_N^mC_{N-m}^m$, with
$%
C_N^m$ being the binomial coefficients. Since the created pairs of droplets
are independent, we write the partition function in the form
\begin{equation}  \label{Z}
Z = \sum_{m=0}^{N/2} P_N^m \exp(-m \beta), \ \ \ \beta = A/k_BT.
\end{equation}
Though the sum can be evaluated exactly and expressed in terms of the
Legendre polynomials for arbitrary $N$, it is more convenient to use the
Stirling formula for the factorials and the condition that the sample is
macroscopic, $N \gg 1$. Approximating the sum by an integral,
\begin{eqnarray*}
Z & = & \int_0^{N/2} \,dm \\
& \times & \exp \left[ -m \beta - N \ln \left( 1-\frac{2m}{N} \right) + 2m
\ln \left( \frac{N}{m}-2 \right) \right],
\end{eqnarray*}
calculating it in the saddle-point approximation, and subsequently
evaluating in the same way the statistical average of $N_2$,
\begin{equation}  \label{N2}
\bar N_2 = Z^{-1} \sum_{m=0}^{N/2} m P_N^m \exp(-m \beta) = -
\frac{\partial%
}{\partial \beta} \ln Z,
\end{equation}
we easily obtain
\begin{eqnarray}  \label{averN2}
\bar N_2 & = & N \exp(-A/2k_BT),  \nonumber \\
\bar N_1 & = & N - 2\bar N_2 = N \left[ 1-2 \exp(-A/2k_BT) \right].
\end{eqnarray}

\begin{figure} \narrowtext
{\epsfxsize=7cm\centerline{\epsfbox{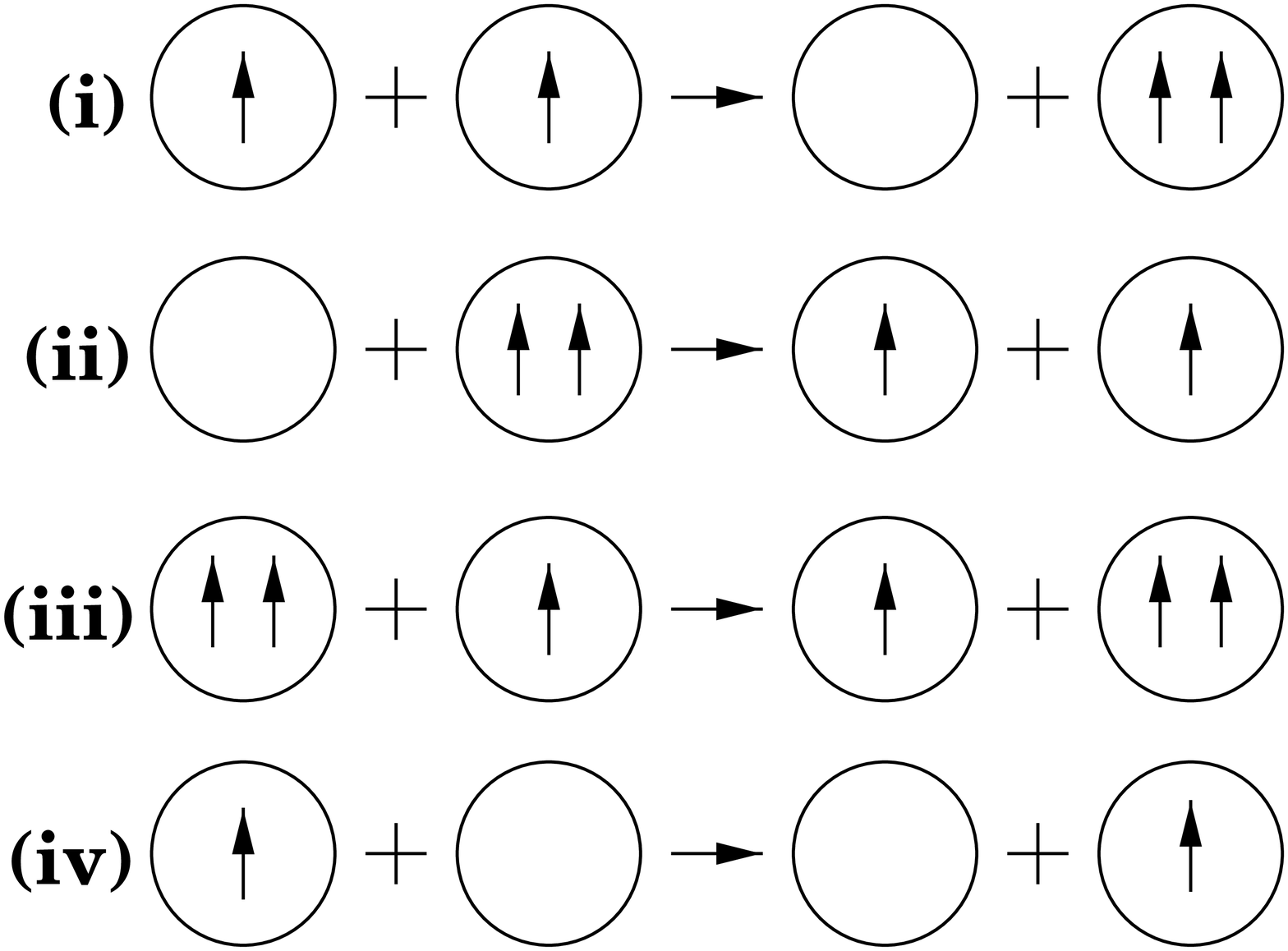}}}%
\caption{Elementary tunneling processes.}
\label{fig1}
\end{figure}

Now we calculate the conductivity. Within the framework of the proposed
model the electron tunneling occurs via one of the four following processes
illustrated in Fig.~1,

\begin{description}
\item  {(i)} In the initial state we have two droplets in the ground state,
and after tunneling in the final state we have an empty droplet and a
droplet with two electrons;

\item  {(ii)} An empty droplet and a two-electron droplet in the initial
state transform into two droplets in the ground state (two droplets with one
electron);

\item  {(iii)} A two-electron droplet and a single-electron droplet exchange
their positions by transferring an electron from one droplet to the other;

\item  {(iv)} An empty droplet and a single-electron droplet exchange their
positions by transferring an electron from one droplet to the other.
\end{description}

In the linear regime, all these processes contribute to the current density
$j$ independently, $j = j_1 + j_2 + j_3 + j_4$. The contributions of the
first two processes read
\begin{equation}  \label{cur1}
j_{1,2} = en_{1,2} \left\langle \sum_i v^i_{1,2} \right\rangle,
\end{equation}
where $n_{1,2} = N_{1,2}/V_s$ are the densities of the single- and
two-electron droplets, and $\langle \dots \rangle$ stands for statistical
and time averages. The appearance of the factors $n_{1,2}$ reflects the fact
that the electron tunnels {\em from} a single-electron droplet (process (i))
or two-electron (ii) droplet. The summation in Eq. (\ref{cur1}) is performed
over all magnetic polarons the electron can tunnel {\em to} --- one-electron
droplets for the process (i) and empty droplets for the process (ii).
Finally, the components of average electron velocity $\langle v^i_{1,2}
\rangle$ along the direction of the electric field are obviously found as
\cite{Mott}
\begin{equation}  \label{veloc1}
\left\langle \sum_i v^i_{1,2} \right\rangle = \left\langle \sum_i \frac{r^i
\cos \theta^i}{\tau_{1,2} (r^i, \theta^i)} \right\rangle,
\end{equation}
where $r^i$ and $\theta^i$ are the electron tunneling length (the distance
between the droplets) and the angle between the electric field and the
direction of motion, respectively, and $\tau_{1,2} (r^i, \theta^i)$ are
characteristic times associated with the tunneling processes. The relation
between $\tau_1 (r,\theta)$ and $\tau_2 (r,\theta)$ can be found from the
following considerations. Near the equilibrium, the number of two-electron
droplets, excited per unit time, equals to the number of the decaying
two-electron droplets. We thus have the detail balance relation,
\begin{equation}  \label{equil}
\frac{\bar N_1^2}{\tau_1 (r, \theta)} = \frac{\bar N_2^2}{\tau_2 (r,
\theta)}%
,
\end{equation}
where we have taken into account that the probability of the formation of a
two-electron droplet is proportional to the total number $N_1$ of the
single-electron states multiplied by the number of available hopping
destinations, which also equals $N_1$. Similarly, the probability of decay
of a two-electron droplet is proportional to $N_2N_3 = N_2^2$. Eq. (\ref
{equil}) implies $\tau_2 (r, \theta) = \tau_1 (r, \theta) \exp(-A/k_BT)$. We
write then the conventional expression for the tunneling times \cite{Mott}
in the following form,
\begin{equation}  \label{tau12}
\tau_{1,2} (r, \theta) = \omega_0^{-1} \exp \left( \frac{r}{l} \pm
\frac{A}{%
2k_BT} - \frac{eEr \cos\theta}{k_BT} \right),
\end{equation}
where $l$ and $\omega_0$ are the characteristic tunneling length and magnon
frequency, and we have taken into account the contribution of the external
electric field to the tunneling probability.

To perform the averaging, we assume that the centers of the magnetic
polarons are randomly positioned in space and the average distance
$n^{-1/3}$
between them is much larger than the droplet radius $a$. Both assumptions
seem to be perfectly justified far below the percolation threshold. Then the
averaged sum in Eq. (\ref{cur1}) is essentially the space average of $v^i$,
multiplied by the number of droplets available for hopping ($N_1$ for the
process (i) and $N_2$ for the process (ii)). Expanding in $eEl/k_BT \ll 1$,
we obtain
\begin{eqnarray}  \label{veloc2}
\left\langle \sum_i v^i_{1,2} \right\rangle & = & \frac{eE\omega_0}{k_BT}
N_{1,2} e^{-A/2k_BT} \left\langle r^2 \cos^2 \theta e^{-r/l}
\right\rangle_V,
\nonumber \\
\bigl\langle \dots \bigr\rangle_V & = & V_S^{-1} \int \dots \ d^3 \bbox{r}.
\end{eqnarray}

In Eq. (\ref{veloc2}), electric field is outside the averaging. Rigorously
speaking, this means that the characteristic hopping length $l$ is
larger than the interdroplet distance $n^{-1/3}$ and our approach is
valid only when the droplet concentration is not too small.

Substituting Eq. (\ref{veloc2}) into Eq. (\ref{cur1}) and performing the
integration, we find
\begin{equation}  \label{cur11}
j_{1,2} = \frac{32\pi e^2 E \omega_0 l^5 n_{1,2}^2}{k_BT} \exp(-A/2k_BT).
\end{equation}

In the processes (iii) and (iv) the free energy of the system is not changed
after the tunneling, and we write the characteristic times as
\begin{equation}  \label{tau3}
\tau_{3,4} (r,\theta) = \omega_0^{-1} \exp(r/l - eEr \cos\theta/k_BT).
\end{equation}
The contribution of these two processes to the current is calculated
similarly to that of (i) and (ii). For the process (iii) the number of
magnetic polarons from which the electron may tunnel, is $N_2$, whereas the
number of accepting droplets is $N_1$. In the same way, for the process (iv)
these numbers are $N_1$ and $N_3 = N_2$, respectively. Consequently, the
factors $n_{1,2}^2$ in Eq. (\ref{cur11}) are replaced by $n_1n_2$,
\begin{equation}  \label{cur31}
j_{3,4} = \frac{32\pi e^2 E \omega_0 l^5 n_1 n_2}{k_BT}.
\end{equation}

From Eqs. (\ref{cur11}) and (\ref{cur31}) we now obtain the dc conductivity
$%
\sigma = j/E$,
\begin{equation}  \label{cond1}
\sigma = \frac{32\pi e^2 \omega_0 l^5}{kT} \left[ 2n_1n_2 + n_1^2
e^{-A/2k_BT} + n_2^2 e^{A/2k_BT} \right].
\end{equation}

In this Section we are only interested in the average conductivity;
fluctuations lead to the appearance of noise and are considered in Section
\ref{noissect}. Using Eq. (\ref{averN2}), we find that all the four
processes illustrated in Fig.~1 give identical contributions to the
conductivity; for $A \gg k_{B}T$ the average conductivity (for which we
retain the notation $\sigma$) reads
\begin{equation}  \label{cond2}
\sigma = \frac{128\pi e^2 n^2 \omega_0 l^5}{k_BT} \exp(-A/2k_BT).
\end{equation}

We see that the conductivity increases with temperature as $\sigma(T)
\propto T^{-1}\exp(-A/2k_BT)$, which is typical for tunneling systems (see
{\em e.g.} Ref. \onlinecite{Mott}).

At this point, let us discuss the applicability range of our model. The
essence of our picture is the existence of different types of droplets. Of
course, only single-electron droplets are stable. Obviously, an empty
droplet decays during the time of the order of $1/ \omega_0$. On the other
hand, following the above discussion, the empty droplet should acquire an
electron from neighboring one-electron or two-electron droplets during the
characteristic time $\tau_0$, which can be easily calculated based on the
following considerations. The probability $P$ per unit time for an empty
droplet to acquire one electron can be written as
\begin{equation}  \label{emptydr}
P = 4 \pi \omega_0 \int_0^{\infty} e^{-r/l}(n_1+n_2 \exp(A/k_BT))r^2 dr,
\end{equation}
where the terms with $n_1$ and $n_2$ correspond to the electron transfer
from single- and two- electron droplets, respectively. Performing
integration in Eq. (\ref{emptydr}) and using Eq. (\ref{averN2}), we find
\begin{equation}  \label{lifetime}
\tau_0 = \frac{\exp(-A/2k_BT)} {8\pi \omega_0 l^3 n}
\end{equation}

Just the same estimate can be obtained for the characteristic time of
electrons leaving two-electron droplets. For our picture with empty and
two-electron droplets to be valid, the following condition must be met, $%
\tau_0 \ll \omega_0^{-1}$. Thus, our approach is valid at sufficiently low
temperatures, $k_BT \ll A$, and for not too small droplet density $n$.

The applicability of our approach also implies that $l > a, n^{-1/3}$. It is
of interest to consider also the case of $l \sim a$ and/or low droplet
concentrations. In this situation, in usual hopping systems, the
conductivity strongly depends on the geometry of current paths \cite{ES}.
This causes exponential dependence of conductivity on the carrier
concentration. However, our system turns out to be more complicated that
those commonly invoked for hopping conductivity. It involves different types
of hopping centers giving rise to unusual geometry of current paths.
Therefore, the conventional approaches used for hopping can not be applied
straightforwardly to the analysis of our model at low droplet concentration
or at $l \sim a$. Despite these complications, we believe that the
expression for the conductivity in the case $l \lesssim n^{-1/3}$ includes
the percolation-related factor $\exp(-\beta/n^{1/3}a)$, with $\beta$ of the
order one \cite{ES}, though currently we have no rigorous proof of this
statement. The results below for magnetoresistance and noise are insensitive
to this factor, and therefore we expect them to be valid in a general case.

\section{Magnetoresistance}

As we already discussed, below the percolation threshold when the
volume fraction of droplets $n < n_c$, a typical value of $A/k_B$ is
mainly determined by Coulomb interactions between two electrons inside
the droplet $A\sim e^2/\epsilon a$ and has a typical value of $1000
K$. Now we can use this estimate to analyze the magnetoresistance in
non-metallic phase-separated manganites. To do that, we use the
expression for the radius of the magnetic polaron, obtained in the
introduction $a \propto d(t/JS^2)^{1/5}$. Recall once more that here
$J \sim$100 K is an AFM Heisenberg exchange between the local spins
$S=3/2$. It is natural to conclude that in the magnetic field $H$ the
Heisenberg exchange integral $J$ decreases according to the formula
$J(H)S^2 = J(0)S^2- g\mu_BHS$, where $\mu_B$ and $g$ are the Bohr
magneton and the gyromagnetic ratio, respectively. Consequently, the
value of $A$ is decreasing linearly in the experimentally accessible
range of magnetic fields, and for the excitation energy we obtain 
\begin{equation}  \label{Curie1}
A(H) = A(0) [1 - bH], \ \ \ b = \frac{1}{5} \frac{g \mu_B}{J(0)S}.
\end{equation}
It follows now from Eq. (\ref{cond2}) that the magnetoresistance is
negative and for temperatures $T < A/k_{B}$ reads
\begin{eqnarray}  \label{magres}
|MR| \equiv \frac{\rho(0)-\rho(H)}{\rho(H)} & = & \exp \left
[\frac{A(0)-A(H)%
}{2k_BT} \right] - 1  \nonumber \\
& = & \exp \left( \frac{bHA}{2k_{B}T} \right) - 1.
\end{eqnarray}
For low magnetic fields and not very small temperatures the absolute
value of the magnetoresistance is small, $|MR| = bHA/2k_{B}T \ll
1$. In higher fields (but still $bH \ll 1$) the absolute value of
magnetoresistance eventually exceeds one and behaves in the
exponential fashion, $|MR|=\exp(bHA/2k_{B}T)$. Note that for
temperatures $T \lesssim A/k_{B}$ and for typical gyromagnetic ratios
$g \sim 10$ the magnetoresistance in our region of doping becomes
larger than one by absolute value only in relatively high magnetic
fields $H \sim 10$T. 

\section{{\mbox{$1/f$}} noise power}

\label{noissect}

Recently, Podzorov {\em et al} \cite{Podzorov} reported the
observation of giant $1/f$ noise in perovskite manganites in the phase
separated regime. Generally, systems with distributed hopping lengths
are standard objects which exhibit $1/f$ noise (for review, see
Refs. \onlinecite{Dutta,Kogan}). The purpose of this Section is to
study low-frequency noise within the framework of the model used to
calculate the conductivity in Section \ref{condsect}, and show that it
has, indeed, $1/f$ form. 

Starting from the Ohm's law $U = IL/\sigma S$ (where $L$ and $S$ are
the sample length and the cross-section, respectively) and assuming
that the measuring circuit is stabilized ($I = const$), we can present
the voltage noise at the frequency $\omega$, $\langle \delta U^2
\rangle_{\omega}$, in the following way, 
\begin{equation}  \label{fluct1}
\langle \delta U^2 \rangle_{\omega} = U_{dc}^2 \frac{\langle \delta \sigma^2
\rangle_{\omega}}{\sigma^2},
\end{equation}
where $U_{dc}$ is the time-averaged voltage, and $\langle \delta
\sigma^2 \rangle_{\omega}$ is the noise spectrum of the fluctuations
of the conductivity. 

If we disregard possible fluctuations of temperature in the system,
the only source of the fluctuations in our model is those of the
occupation numbers $n_1$ and $n_2$. Using the conservation law $n_1 +
2 n_2 = n$, we find from Eq. (\ref{cond1}) 
\begin{equation}  \label{condfl1}
\delta \sigma = \sigma \frac{\delta n_2}{\bar n_2} \left[ 1 -
2\exp(-A/2k_BT) \right].
\end{equation}
We thus need to find the fluctuation spectrum $\langle \delta n_2^2 
\rangle_{\omega}$. Following the general prescription \cite{LL}, we
recollect that the two-electron droplets decay via the process (ii),
and the relaxation equation has the form
\begin{equation}  \label{kineq1}
\delta \dot n_2 = -\frac{\delta n_2}{\tau(r)}, \ \ \ \tau (r) =
\omega_0^{-1} \exp(r/l - A/2k_BT),
\end{equation}
where we have neglected the effect of the electric field. The
fluctuation spectrum then reads \cite{LL}
\begin{eqnarray}  \label{nfl1}
\langle \delta n_2^2 \rangle_{\omega} & = & \langle \delta n_2^2 \rangle_{T}
\left\langle \sum_i \frac{2\tau(r^i)}{1 + \omega^2\tau^2 (r^i)}
\right\rangle
\nonumber \\
& = & 8 \pi \bar n_2 \langle \delta n_2^2 \rangle_{T} \int_0^{\infty}
\frac{%
\tau(r)}{1 + \omega^2\tau^2 (r)} r^2 dr,
\end{eqnarray}
where $\langle \delta n_2^2 \rangle_T$ is the thermal average of the
variation of $n_2$, and the summation is performed over the pairs "empty
droplet -- two-electron droplet", with $r_i$ being the distance
between the sites in a pair. The average in Eq. (\ref{nfl1}) is
essentially a spatial integral, with the main contribution coming from
short distances, 
\begin{equation}  \label{nfl2}
\langle \delta n_2^2 \rangle_{\omega} = 8 \pi \bar n_2 \langle \delta n_2^2
\rangle_{T} \int_0^{\infty} \frac{\tau(r)}{1 + \omega^2\tau^2 (r)} r^2 dr.
\end{equation}
Note that Eq.(\ref{nfl2}) is valid for an arbitrary relation between
$l$ and $a$, not necessarily for $a \ll l$. 

We are interested below in the frequency range
\begin{equation}  \label{freqr}
\tilde \omega_0 \exp (- L_s/l) \ll \omega \ll \tilde \omega_0, \ \ \ \tilde
\omega_0 \equiv \omega_0 \exp (A/2k_BT),
\end{equation}
where $L_s$ is the smallest of the sample sizes. In this case, with
the logarithmic accuracy we obtain for $A \gg k_{B}T$,
\begin{eqnarray}  \label{nfl3}
\langle \delta U^2 \rangle_{\omega} & = & U_{dc}^2 \frac{\langle \delta
n_2^2 \rangle_{T}}{\bar n_2} \frac{4\pi^2 l^3}{\omega} \ln^2 \left(\frac{%
\tilde\omega_0}{\omega} \right).
\end{eqnarray}
Thus, in the wide range of sufficiently low frequencies (\ref{freqr})
the noise power spectrum for our system has almost $1/f$ form.

The variation $\langle \delta n_2^2 \rangle_T = V_S^{-2}
(\overline{N_2^2} - \bar N_2^2)$ is easily found in the same way as
Eq. (\ref{averN2}), 
\begin{equation}  \label{nfl5}
\left\langle \delta n_2^2 \right\rangle_T = \frac{\bar n_2}{2V_s}.
\end{equation}
Combining this with Eq. (\ref{nfl3}), we write the final expression
for the spectral density of noise for $A \gg k_{B}T$ in the form
\begin{eqnarray}  \label{voltfl1}
\langle \delta U^2 \rangle_{\omega} & = & U_{dc}^2 \frac{2\pi^2 l^3}{V_s
\omega} \ln^2 ( \frac{\omega_0 e^{A/2k_BT}}{\omega}).
\end{eqnarray}

\section{Discussion}

\label{discuss}

For the further discussion, it is convenient to rewrite
Eq. (\ref{voltfl1}) in the form
\begin{equation}  \label{voltfl2}
\alpha = \frac{\langle \delta U^2 \rangle_{\omega} V_s \omega}{U^2_{dc}} =
2\pi^2 l^3 \ln^2 (\frac{\tilde \omega_0}{\omega}).
\end{equation}

It is remarkable that the noise spectrum in our model has $1/f$ form up to
very low frequencies. This is due to fluctuations in occupation numbers of
droplets, associated with creation and annihilation of extra electron --
hole pairs. This mechanism of $1/f$ noise is specific for our model and is
not present in standard hopping conduction.

Let us estimate the numerical value of the parameter $\alpha$, which
is the standard measure of the strength of $1/f$ noise. This parameter
is proportional to the third power of $l$. Simple estimates (analogous
to that presented in the Introduction for $a$) reveal that, in
general, $l$ of the order or higher than $a$. Assuming again that the
excitation energy is of the order of the Coulomb energy $A \sim
e^{2}/a \epsilon$, taking $\omega_0$ to be of the order of the Fermi
energy inside droplets (which means $\hbar\omega_0 \sim 300 K$ for $n
< n_c$), and estimating the tunneling length $l$ as being $l \gtrsim
2a \sim$20 \AA, we arrive to a conclusion that the parameter $\alpha $
is of the order $\alpha \approx 10^{-17}$--$10^{-16}$cm$^3$ for $T <
A/k_{B}$ and $\omega \sim $1 Hz -- 1 MHz. This value of $\alpha$ is by
several orders of magnitude higher than in usual semiconducting
materials (see Ref. \onlinecite{Dutta}). Such a large magnitude of the
noise can be attributed to the relatively low height of the potential
barrier $A$ and to the relatively large tunneling length
$l$. Formally, it is also related to the large value of the logarithm
squared in Eq. (\ref{voltfl2}). 

According to Eqs. (\ref{voltfl1}) and (\ref{voltfl2}), the noise power and
the noise parameter $\alpha$ are independent of the volume fraction occupied
by the droplets. This result is valid in the intermediate range of $n$, when
the droplet density is not too high and not too low. First, we assumed that
the droplets are isolated point objects and the tunneling between the two
droplets is not affected by a third polaron. This is only valid provided the
droplet density is far from the percolation threshold, $n \ll n_c$. On the
other hand, the droplet density must not be too low since the conditions $N,
N_1, N_2 \gg 1$ are assumed to be met. Moreover, we neglected the
possibilities of the disappearance of a droplet without an electron, the
formation of a new droplet due to the electron tunneling, and the decay of
two-electron droplets. Thus, the characteristic times of these processes
should be longer than the characteristic tunneling time, and the average
tunneling distance can not be too high (see Eq. (\ref{lifetime}) and the
discussion below it).

The above speculations imply that the following set of inequalities should
be met, $a \ll n^{-1/3} \ll l$, for formula (\ref{cond2}) for the
conductivity to be valid. In general, the tunneling length should not be
much larger than the droplet radius since just the same physical parameters
determine these two characteristic distances. So, these inequalities could
not be valid for real physical systems and it is of interest to consider the
situation where $a, l \ll n^{-1/3}$, which is beyond the scope of our model.
However, some definite conclusions concerning the magnetoresistance and the
noise power can be made at present.

First, the factor $\exp(A/2k_BT)$ in the temperature dependence of the
conductivity is related to the number of carriers and appears due to
strong Coulomb repulsion of electrons in the droplet. It seems rather
obvious that such a factor appears in the formula for the conductivity
below the percolation threshold for an arbitrary relation between $a$
and $l$. On the other hand, in contrast to common hopping systems,
strong $1/f$ noise in our model results from fluctuations of state
occupation numbers. Actually, our result for parameter $\alpha$
(\ref{voltfl2}) only relies on the fact that $\delta \sigma/\sigma
\sim \delta n_2/\bar n_2$, and thus the noise power is expected to
have $1/f$ form below the percolation threshold for any relation
between $a$ and $l$. Then, we can also conclude that in our model
strong $1/f$ noise will be found under the (experimentally relevant)
conditions $a \sim l$ .

Another important point is that we disregard direct Coulomb
interaction between the droplets in comparison with the energy
$A$. This is justified if the gas of the droplets is diluted,
$n^{-1/3} \gg a$. In this respect, we recollect that in standard
hopping conduction systems (doped semiconductors) the main mechanism
of low-frequency noise is an exchange of electrons between the
infinite cluster and nearby finite clusters. In the absence of
interactions it leads to the noise power proportional to
$\omega^{-\alpha}$, with the exponent $\alpha$ being considerably
below one \cite{KS}. To explain $1/f$ noise in these systems, models
involving Coulomb interactions have been proposed
\cite{Kozub,Yu}. These sources of low-frequency noise are thus 
beyond our discussion. We also did not consider sources of noise
different from resistance fluctuations. At least two other types of
noise are inevitably present in the system, Nyquist-Johnson (thermal)
noise, which is a consequence of the fluctuation-dissipation theorem,
and shot noise due to the discrete nature of electron charge (see
Refs. \cite{Kogan,BB00} for review). Both these noises are frequency
independent (white) at low frequencies. The magnitudes of
Nyquist-Johnson, shot, and $1/f$ noises are governed by absolutely
different parameters, and we do not attempt to compare them here,
noting only that at low frequencies $1/f$ noise must dominate.

In our model, we assumed that the number of droplets $N$ is fixed and
strictly equal to the number of extra electrons. In actual systems,
$N$ can also fluctuate, and this can be an additional source of noise,
and of $1/f$ noise, in particular. However, this contribution depends
critically on the heights of corresponding energy barriers and can
vary for different systems. 

As we have already mentioned, the main motivation of our work was the
experimental study \cite{Podzorov}, which observed high $1/f$ noise
power at high temperatures far from the metal-insulator transition. In
the same experiment, noise dropped to much lower levels at low
temperatures in the metallic phase. This behavior of the noise power
is consistent with the present model since in the metallic phase the
electron tunneling contribution to the total conductivity is
negligible. In the vicinity of the percolation transition the noise
power increases drastically \cite{Podzorov}. In this paper we do not
attempt to describe the system of magnetic polarons close to the
percolation threshold. However, we argue that the amplitude of $1/f$
noise is already large in the phase-separated regime even far from the
percolation threshold. 

\section{Conclusions}

We emphasize that even in our oversimplified model we get a reasonable
behavior of resistivity and magnetoresistance for underdoped manganites.
Moreover, we have shown that in the framework of our model $1/f$ noise
appears in the natural way. The phase-separation ensures a large magnitude
of the noise power as compared with homogeneous materials.

Of course, a more sophisticated theory should include both the
ferromagnetic structure of the droplet and the antiferromagnetic
structure of the insulating matrix. This can lead us to the physics
resembling that observed in the process of spin-assisted tunneling,
which attracts a considerable interest nowadays (see {\em e.g.}
Ref. \onlinecite{spin-tunnel}). The work in this direction is in
progress. 

\section*{Acknowledgments}

This work was supported by Swiss National Science Foundation, INTAS (grants
97-0963 and 97-11954), Russian Foundation for Basic Research (grants
00-02-16255 and 00-15-96570), and the Russian President Program (grant
96-15-9694). We acknowledge stimulating discussions with P.~Fulde,
D.~Vollhardt, K.~Held, P.~W\"{o}lfle, A.~F.~Andreev, Yu.~Kagan,
A.~S.~Ioselevich, M.~S.~Mar'enko and D.~V.~Efremov. M.~Yu.~K. would like
also to acknowledge hospitality of the Max-Planck-Institut in Dresden where
a part of this work was done.

\end{document}